# Fourier pixels for reciprocal light control

Yannik M. Glauser[1*], Sander J. W. Vonk[1*], David B. Seda[1], Hannah Niese[1], Boris de Jong[1], Matthieu F. Bidaut[1], Daniel Petter[1], Gabriel Nagamine[1,2], Nolan Lassaline[1,3], and David J. Norris[1]

[1]Optical Materials Engineering Laboratory, Department of Mechanical and Process Engineering, ETH Zurich, 8092 Zurich, Switzerland.
[2]Institute of Applied Physics, University of Bern, 3012 Bern, Switzerland.
[3]Department of Physics, Technical University of Denmark, 2800 Kongens Lyngby, Denmark.

* These authors contributed equally to this work.

**Digital cameras[1] and displays[2] utilise picture elements (pixels[3]) that perform a single function: detecting or emitting light intensity. To exploit the full information content of electromagnetic waves, more advanced elements are required. This has driven the development of multifunctional components, which for example, simultaneously detect and emit intensity[4,5] or extract intensity and spectral information[6-8]. However, no pixel exists that both senses and generates optical wavefronts with full control over amplitude, phase, and polarisation, limiting reciprocal control and feedback of sophisticated light fields. Here we present a route to such pixels by demonstrating a versatile platform of miniaturised diffractive elements based on Fourier optics[9]. We exploit plasmonic surface waves[10], which propagate coherently[11] and efficiently[12-15] across metallic surfaces. When these plasmons are launched towards wavy microstructures[16] designed with simple Fourier analysis, arbitrary and background-free optical wavefronts are generated. Conversely, incoming light can be sensed and its amplitude, phase, and polarisation fully characterised. By combining or superposing several such components, we create multifunctional 'Fourier pixels' that provide compact and accurate control over the optical field. Our approach, which could also use photonic waveguide modes, establishes a scalable, universal architecture for vectorially programmable pixels with applications in adaptive optics[17,18], holographic displays[19-21], optical communication[22,23], and quantum-information processing[24].**

Cameras[1] and displays[2] rely on pixels that locally measure or emit light intensity. Similarly, spatial light modulators[25] contain elements that modulate the optical phase. Yet, these devices modify only one property of the electromagnetic field. To simultaneously control amplitude, phase, and polarisation, multiple bulky elements have traditionally been required. Metasurfaces[26]—thin components constructed from many nanoscale resonators that collectively shape or analyse optical wavefronts—have enhanced our capabilities, leading to polarisation cameras[27], phase detectors[28], phase modulators[29,30], and various optical sources[21,31-35]. However, because sensing and generation require different strategies, metasurface-based pixels that provide both detection and emission of arbitrary wavefronts have not been possible. If available, they would enable reciprocal control and feedback of light with compact, modular devices.

To create such multifunctional pixels, a versatile platform is needed that also utilises a simple inverse-design methodology (to predict the optimal pixel structure for a specific task). Inverse design of metasurfaces is challenging, requiring computationally intensive simulations to select the proper set of discrete resonators and their placement[36]. We pursue a different approach, based on Fourier optics[9]. Diffraction is governed by the spatial frequencies (sinusoids) in a surface profile. Thus, 'wavy' interfaces are well suited to control electromagnetic waves. Further, their inverse design involves Fourier analysis, which is not only straightforward but allows surface profiles for different tasks to be superposed to obtain multifunctionality. However, this approach has been impeded by the inability to fabricate diffractive interfaces containing precise sums of sinusoids. Recently, such 'Fourier surfaces' have become possible[16], providing a potential route to a broad class of multifunctional pixels.

Here we exploit this approach to create 'Fourier pixels', which allow control and detection of optical fields with arbitrary amplitude, phase, and polarisation. Each Fourier pixel uses surface plasmon polariton waves[10] (SPPs), which are launched from a source and then propagate coherently[11] and efficiently[13] across a metallic interface (Fig. 1a). When they encounter a diffractive Fourier element, a precise, background-free optical field is outcoupled. For light shaping, this output is the desired optical wavefront emitted by the pixel. Alternatively, if the SPPs are generated at the source by an incoming optical wave, the output from the Fourier element can fully characterise this incoming light, with the SPPs as intermediaries. Moreover, several diffractive surface profiles can be superposed within the same Fourier element to obtain multifunctionality. Thus, this simple architecture provides a thin, modular, scalable, and universal platform for vectorially programmable reciprocal elements that both shape and sense light fields within footprints comparable to conventional pixels.

As a first example, we demonstrate Fourier pixels that generate light with arbitrary amplitude and phase. We use silver (Ag) surfaces to maximise SPP propagation[13,37] (Methods). The pixel in Fig. 1b creates a Gaussian beam with a twisted phase front that varies clockwise from 0 to $2\pi$ around the beam axis in the direction of propagation (known as a vortex beam with a topological charge, $q$, of $+1$). For simplicity, a sinusoidal grating (left) is employed as the SPP source. Light incident with a specific energy and angle launches SPPs efficiently (~70–80%, Methods). These SPPs then propagate in $x$ towards a diffractive Fourier element (right), which modifies the SPPs in three ways. It imprints the desired phase and amplitude profile on the outcoupled light and corrects for the momentum mismatch between the SPPs and the emitted photons. Despite this complexity, the required design (Fig. 1b, inset) is easily determined by taking the inverse Fourier transform of the targeted complex-valued optical wavefront (amplitude and phase) while anticipating the required SPP-to-photon momentum change (Methods). Figure 1c shows a scanning-electron micrograph (SEM) of the fabricated Fourier pixel. (SEMs and designs for devices not shown in the main figures are presented in Extended Data Fig. 1.)

In Figure 1d, we image the pixel output at the back focal plane (Fourier plane) of an optical microscope (Methods and Extended Data Fig. 2). Due to the phase singularity in the centre of the vortex beam, a characteristic 'doughnut' profile appears.[38] Using a cylindrical lens, which yields an interference pattern (inset) with $|q|$ nodes and a tilt (right or left) that reveals the vortex handedness,[39] we verify the topological charge ($q = +1$) of the beam. Figure 1e,f shows similar results for Fourier pixels that generate vortex beams with higher twist ($q = +3$ and $+5$). To further quantify the quality of the output, Fig. 1g,h presents Fourier pixels (holograms) that project single- or multi-colour flat-phase images in Fourier space. The images exhibit uniform intensity distributions with low speckle contrasts (ratio of standard deviation to mean intensity) of 17%, confirming the high fidelity of the imprinted phase. Further examples of Fourier pixels that generate Hermite–Gaussian beams or that are driven with incoherent light are shown in Extended Data Fig. 3.

The Fourier pixels can also easily incorporate lensing functionality to place the desired output at arbitrary planes (not just in the Fourier plane). For example, by directly encoding a wavy Fresnel-type phase profile on a Fourier element, we create a single focal spot 25 µm above the pixel (Fig. 1i). The measured intensity distribution reveals a diffraction-limited focus (0.27 µm), indicating minimal aberrations. Such pixels also confirm high throughput efficiency, measuring >40% power-in to power-out for wavelengths from 500 to 700 nm (Methods). Extending this concept further, Fig. 1j shows a pixel that generates a grid of foci. Additional results for phase and amplitude control at arbitrary planes, including lenses with various numerical apertures up to 0.8, appear in Extended Data Fig. 4.

However, full control of the emitted optical field requires manipulation of the polarisation. Figure 2a shows a 'vectorial' Fourier pixel that utilises two orthogonal gratings to launch SPPs with distinct wavevectors and in-plane polarisations. By coherently superposing the contributions from these two gratings, the Fourier element generates optical wavefronts with arbitrary amplitude, phase, and polarisation, enabling deterministic construction of emitted light fields. As an example, we launch SPPs with incident light polarised 45° with respect to each of the two orthogonal gratings (for equal amplitude in the two SPP paths) and create vector beams with polarisation singularities. At the output (Fourier) plane, the field is constructed from the SPPs that originated from the $x$ and $y$ polarisation projections of the two gratings (Fig. 2b). Figure 2c,d shows the output from Fourier pixels that generate first- and second-order vector beams with the polarisation rotating $2\pi$ (see Fig.2b) and $4\pi$ clockwise around the beam axis in the direction of propagation, respectively. The beam profile with its vanishing central intensity and its evolution when a linear polariser (analyser) is rotated in the detection path confirms that the generated fields possess the targeted spatially varying polarisation.

The ability to control polarisation in the output enables multiplexing (Fig. 2e,f). By encoding two distinct images into orthogonal polarisation states, a Fourier pixel can generate fields with different polarisation using the same diffractive Fourier element. Rotation of an analyser in the detection path selectively reveals one image while suppressing the other, establishing polarisation-multiplexed image construction with negligible crosstalk.

In addition to light shaping, the same concepts can be exploited for sensing. Figure 3a illustrates a Fourier pixel that measures the phase difference $\varphi_{\text{in}}$ between light incident on two opposing sinusoidal gratings. These gratings launch counter-propagating SPPs towards a central Fourier element, designed to diffract the two outcoupled light fields together. To determine $\varphi_{\text{in}}$ without a reference wave, the Fourier element also adds a phase difference $\varphi_{\text{out}}$ between the two light fields that depends on the outcoupled angle $\theta$. $\varphi_{\text{out}}$ varies from 0 to $2\pi$ and back to 0 over a range of $\theta$ (along $x$). This additional phase $\varphi_{\text{out}}(\theta)$ (Fig. 3a, inset) results in constructive and destructive interference between the output waves as a function of $\theta$. Consequently, the output (Fourier) plane (Fig. 3b) reveals two nodes arising from destructive interference at a specific angle $\theta$. The Fourier element is also designed to outcouple light over a range of transverse momenta $k_y$ that share the same $\varphi_\text{out}$. This redundancy facilitates coherent overlap of the two outcoupled fields, improving signal visibility and robustness against fabrication or alignment imperfections.

Figure 3c shows the optical output versus $\varphi_\text{out}$, with $\hat{\varphi}_\text{out}$ marking the output phase with maximal destructive interference (at two outcoupling angles $\theta$). The measured phase difference between the input beams is $\hat{\varphi}_\text{in} = \pi - \hat{\varphi}_\text{out}$. By systematically varying the input phase difference $\varphi_\text{in}$ with a spatial-light modulator (Methods), we retrieved $\hat{\varphi}_\text{in}$ from the measured $\hat{\varphi}_\text{out}$ (Fig. 3d), finding excellent agreement. Indeed, the Fourier pixel was so sensitive that it was limited by our experimental setup. Namely, changes in the pixel output over time were consistent with fluctuations of 0.07° in the laser that launched the SPPs (Extended Data Fig. 5). In Extended Data Fig. 6, we also demonstrate an alternative phase sensor that exploits SPP interference directly in the pixel plane.

For polarisation detection, we designed and fabricated a vectorial Fourier pixel that combines the functionalities of linear polarisers, half-wave plates, and quarter-wave plates to provide a full Stokes sensor in a single thin element. Light with unknown polarisation $\mathbf{E}_\text{in}$ strikes two orthogonal gratings (A and B, as in Fig. 2a, but with a different Fourier element). The SPPs launched in the two paths contain information about the amount of $x$ and $y$ polarisation in $\mathbf{E}_\text{in}$ as well as their relative phase. The Fourier element then overlays two images at the output (Fourier) plane. SPPs traveling in $x$ generate a flat-amplitude, rectangular image with no added phase $\varphi_\text{out}$ (Fig. 3e), while SPPs traveling in $y$ generate an orthogonal rectangular image subdivided into regions in which the Fourier element adds $\varphi_\text{out}$ of 0, $\pi/2$, $\pi$, and $3\pi/2$ (Fig. 3f). When these two rectangles are coherently superposed, the resulting output field is

$$\mathbf{E}_\text{out} = \mathbf{E}_\text{in} \cdot \hat{\mathbf{x}} + (\mathbf{E}_\text{in} \cdot \hat{\mathbf{y}}) e^{i\varphi_\text{out}}.$$

Thus, by reading the intensity in the different regions in the output plane (Fig. 3g) with a fixed analyser, the Stokes parameters ($S_0$, $S_1$, $S_2$, and $S_3$) for the incident polarisation state are retrieved (Methods). Specifically, the total input intensity ($S_0$) is obtained by summing the orange regions, while the degree of linear polarisation along $x$ or $y$ ($S_1$) is their difference. The two brown central regions compare $0$ and $\pi$ phase shifts added to the $y$ electric field, analogous to inserting a half-wave plate. Their difference provides the degree of linear polarisation along axes ±45° from $x$ ($S_2$). Similarly, the difference between the two green regions, which introduce $\pi/2$ and $3\pi/2$ phase shifts, reveals the degree of circular polarisation ($S_3$).

We tested our Stokes sensor using linearly and circularly polarised input beams. Figure 3h shows the output (Fourier) plane for an input beam with diagonal linear polarisation (see Extended Data Fig. 7 for circular input polarisations). Such raw images represent the experimental version of Fig. 3g. Figure 3i–k shows the retrieved Stokes parameters for diagonal linear ($S_1 = 1$), left-circular ($S_3 = -1$), and right-circular ($S_3 = 1$) polarisation: $S_1 = 0.87$, $S_3 = -0.83$, and $S_3 = +0.77$, respectively. Again, the Fourier pixel was extremely sensitive, with the measured values limited by our inability to generate a sufficiently accurate polarisation state at the pixel due to optical aberrations or slight misalignments.

An additional advantage of Fourier pixels is their ability to easily combine the above capabilities. Multifunctionality is missing in other approaches, preventing compact, single-shot characterisation of complex optical fields. We demonstrate two different approaches to sense amplitude, phase, and polarisation in a single pixel. First, we cluster two phase sensors (for $x$ and $y$) and a Stokes sensor (Fig. 4a), exploiting three Fourier elements that project to different locations in the output (Fourier) plane (Fig. 4b, reconstructed from three separate measurements). Each phase sensor generates a pair of redundant interference bands (as in Fig. 3b) at the edge (top/bottom for $x$ and left/right for $y$); the Stokes sensor generates an image (the same data shown in Fig. 3h) in the centre. By analysing the combined output, the light striking the pixel can be fully analysed. Second, the three individual Fourier elements in Fig. 4a can be superposed (Fig. 4c, with minor design adjustments to the Stokes sensor). Figure 4d,e shows SEMs of two examples: four gratings surround a Fourier element 30×30 and 10×10 µm$^2$ in size, respectively. The outputs (Fig. 4f,g) again provide amplitude, phase, and polarisation detection, but now for a pixel footprint comparable to conventional pixels. We use the pixel in Fig. 4a to map the phase or polarisation of larger complex field profiles (Methods), including optical vortices (Fig. 4h) and trefoil aberrations (Fig. 4i), as well as polarisation singularities of vector beams (Fig. 4j).

We have introduced Fourier pixels based on plasmonic interfaces with shallow (sub-wavelength) wavy patterns to efficiently generate and sense optical fields with arbitrary control over amplitude, phase, and polarisation. The pixels exploit linear diffraction theory, greatly simplifying their design. Although here we used gratings as SPP sources, subwavelength slits or apertures could launch SPPs in optical transmission[40]. Coherent SPPs could also be generated thermally[11], electrically [41], or via energy transfer from quantum emitters[42]. Arrays of Fourier pixels could enable camera–displays for simultaneous detection and generation of light. The Fourier-pixel concept is also transferable to other guided modes in photonic platforms such as silicon-on-insulator, silicon, silicon nitride, or lithium niobate. Finally, the pixel outputs could be adjusted dynamically by the addition of tunable[43,44] or nonlinear materials. Thus, Fourier pixels provide previously unavailable capabilities for reciprocal wavefront control and establish a universal design principle for a broad class of simple-to-design modular optical devices.

**Online Content** Methods, along with any Extended Data display items, are available in the online version of the paper; references unique to these sections appear only in the online paper.

**Acknowledgements** We thank R. Keitel and L. Novotny for stimulating discussions. S.J.W.V. acknowledges support from the Swiss National Science Foundation (SNSF) under Award No. 200021-232257. M.F.B. acknowledges support provided by a fellowship from the German Academic Exchange Service (DAAD). N.L. acknowledges funding from the Swiss National Science Foundation (Postdoc Mobility P500PT_211105) and the Villum Foundation (Villum Experiment 50355).

**Author Contributions** Y.M.G., S.J.W.V., N.L., and D.J.N. conceived the project. Y.M.G. and S.J.W.V. designed the Fourier pixels with input from N.L. and D.J.N. Y.M.G., S.J.W.V., and D.B.S. fabricated the pixel patterns with assistance from H.N. and B.d.J. Y.M.G. and S.J.W.V. performed and analysed the optical measurements with assistance from B.d.J., M.F.B., D.P., and G.N. Y.M.G. and S.J.W.V. developed the analytical model. Y.M.G., S.J.W.V., and D.J.N. wrote the manuscript with input from all authors. D.J.N. supervised the project.

**Author Information** The authors declare the following potential competing financial interests: Y.M.G., S.J.W.V., and D.J.N. have filed a patent application related to ideas in this work. Readers are welcome to comment on the online version of the paper. Correspondence and requests for materials should be addressed to D.J.N. (dnorris@ethz.ch).

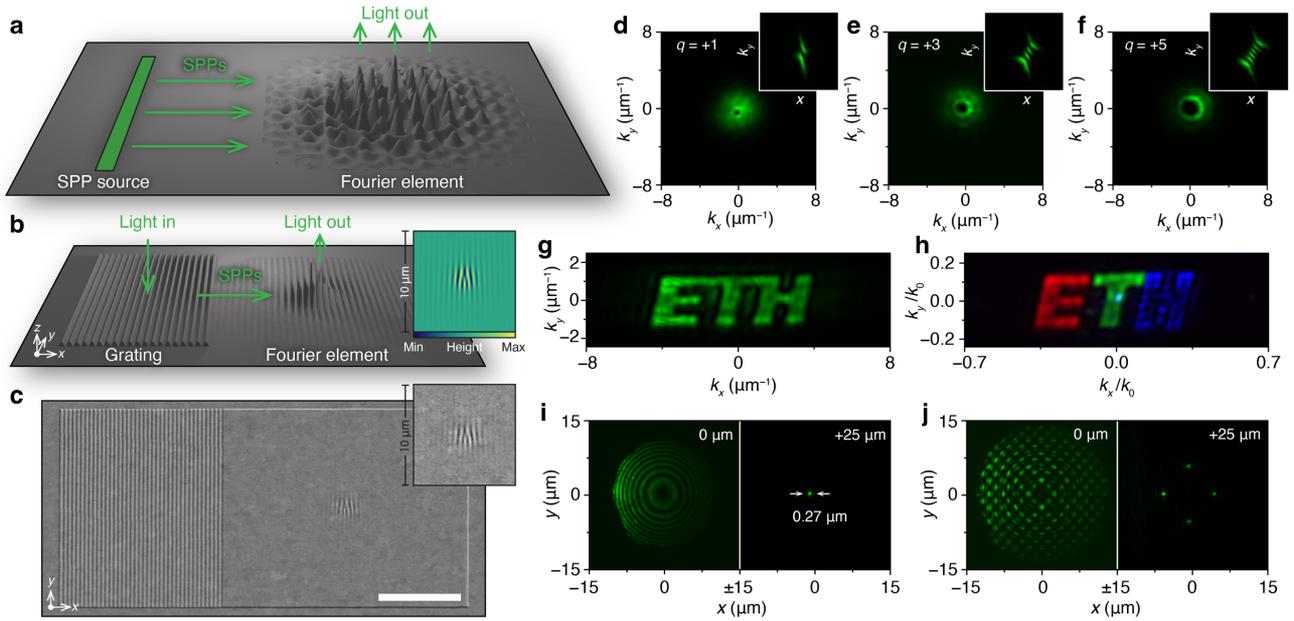

**Figure 1 | Fourier pixels for generating light with arbitrary amplitude and phase. a**, Basic concept of a Fourier pixel, including a source of surface plasmon polaritons (SPPs), a SPP propagation path, and a diffractive Fourier element that generates an optical output. **b**, Fourier pixel that generates a Gaussian vortex beam with a topological charge of $q = +1$. Light generates SPPs at the grating, which launches them towards the Fourier element. The dimensions of the pattern are exaggerated for clarity. The inset shows the surface profile (top-view of the central region) required for the desired output. This pattern simultaneously imprints an amplitude and phase profile on the outcoupled light while correcting for the momentum mismatch between the SPPs and emitted photons. **c**, Scanning electron micrograph (SEM) of the structure depicted in **b** and its inset (35° tilt). The scale bar is 10 μm. **d**, Gaussian vortex beam with $q = +1$ from the Fourier pixel in **c** imaged at the back focal plane (Fourier plane) of an optical microscope. While the dark hole in the centre of the beam profile is indicative of the phase singularity, we independently verify the value of $q$ by inserting a cylindrical lens in the beam path (inset). **e,f**, Gaussian vortex beams imaged as in **d** with $q = +3$ and $+5$, respectively. **g,h**, Experimental single- and multi-colour flat-phase holograms in Fourier space. The uniformity of the images indicates high fidelity of the phase profile. Multi-colour operation is achieved by using three sinusoids to launch SPPs at multiple wavelengths with the same incident angle and the image is reconstructed using three separate measurements for different colours. **i,j**, Fourier pixels that generate a single diffraction-limited focal spot (standard deviation of 0.27 μm) or a grid of foci 25 μm above the Fourier pixel. In each case, a pair of images are shown. The microscope is focused at the pixel plane (left; $z = 0$ μm) and 25 μm above the pixel plane (right; $z = 25$ μm).

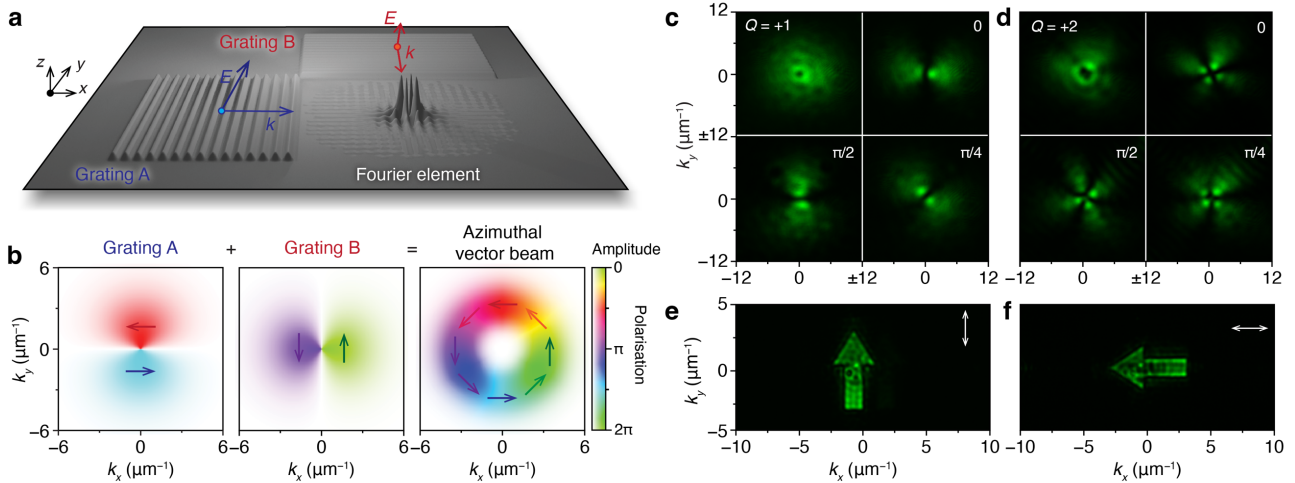

**Figure 2 | Vectorial Fourier pixels for full light control and multiplexing. a,b,** Vectorial Fourier pixel consisting of two orthogonal gratings. Gratings A and B launch SPPs with orthogonal wavevectors and in-plane polarisations. The input waves from A and B can be coherently superposed to generate light fields with arbitrary amplitude, phase, and polarisation. In **b**, the construction of an azimuthally polarised vector beam is depicted. **c,d,** Experimental images of first- ($Q = +1$) and second-order ($Q = +2$) vector beams in Fourier space without a linear polariser in the detection path (top left) and with varying polariser angles between 0 and $\pi/2$. The polarisation singularity introduces a dark hole in the vector beam. **e,f,** Polarisation-multiplexed images of upward and sideward arrows. The separate images are selected by a linear polariser (white arrow orientation) in the detection path.

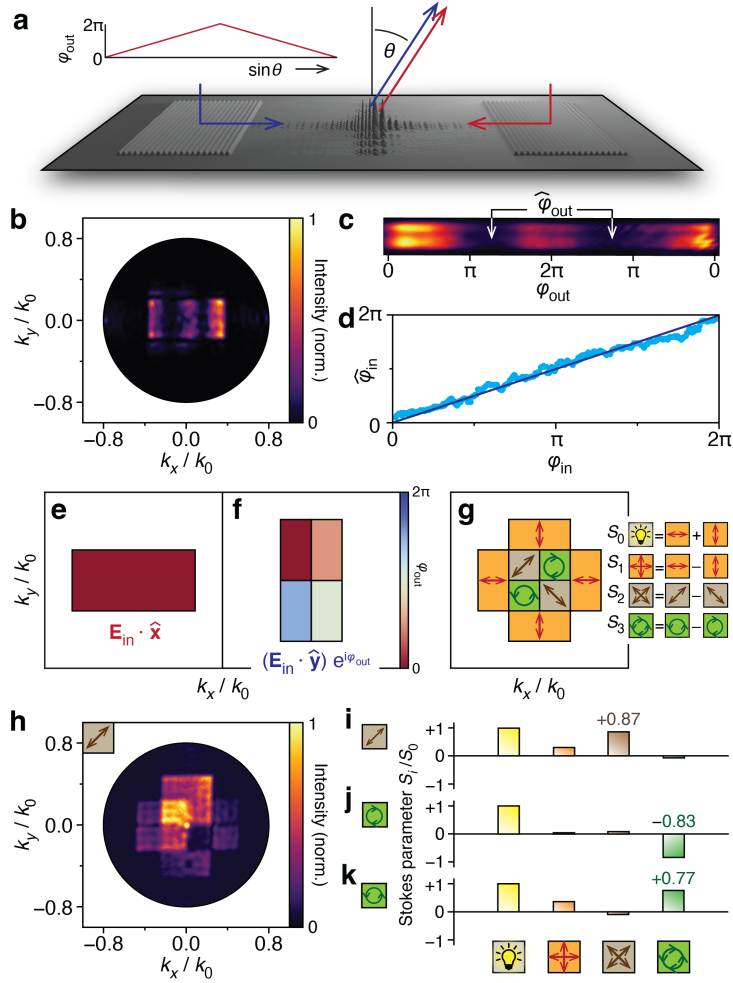

**Figure 3 | Fourier pixels for amplitude, phase, and polarisation sensing. a**, Schematic of a Fourier pixel that measures the phase difference $\varphi_{\text{in}}$ between light incident on two opposing sinusoidal gratings. The light launches counter-propagating SPPs towards a central diffractive Fourier element. This element adds a phase difference $\varphi_{\text{out}}$ between the two diffracted SPP fields that depends on the output angle $\theta$ (inset). This leads to constructive and destructive interference in the output plane (Fourier space). **b**, Measured optical output for an input phase difference $\varphi_{\text{in}} = 0$, showing two dark nodes corresponding to destructive interference at specific normalised in-plane momenta $k_x/k_0$. **c**, The optical output, highlighting the definition of the output phase difference $\varphi_{\text{out}}$ and the interference minima at $\hat{\varphi}_{\text{out}}$ used to determine the measured phase difference $\hat{\varphi}_{\text{in}} = \pi - \hat{\varphi}_{\text{out}}$. **d**, Experimental calibration of the sensor shows excellent agreement between the extracted $\hat{\varphi}_{\text{in}}$ and applied phase differences $\varphi_{\text{in}}$. **e,f**, Concept of a vectorial Fourier pixel functioning as a full Stokes sensor (amplitude and polarisation) by introducing controlled phase retardations between two orthogonal SPP modes launched from two orthogonal gratings (as in Fig. 2a). **e**, SPPs traveling in $x$ generate a rectangular image at the output plane in Fourier space with flat amplitude and no added phase. **f**, SPPs traveling in $y$ generate an orthogonal flat-amplitude rectangular image subdivided into regions in which the Fourier element adds $\varphi_{\text{out}}$ of $0$, $\pi/2$, $\pi$, and $3\pi/2$ with respect to the rectangular image from $x$. **g**, The output intensities in different regions of the overlaid rectangular images yield the Stokes parameters $S_0$, $S_1$, $S_2$, and $S_3$ as shown, enabling retrieval of the total intensity and the degree of linear/circular polarisation (see Methods). **h**, Measured optical output for an input with diagonal linear polarisation. **i–k**, Retrieved Stokes parameters for inputs with diagonal linear, left-circular, and right-circular polarisation.

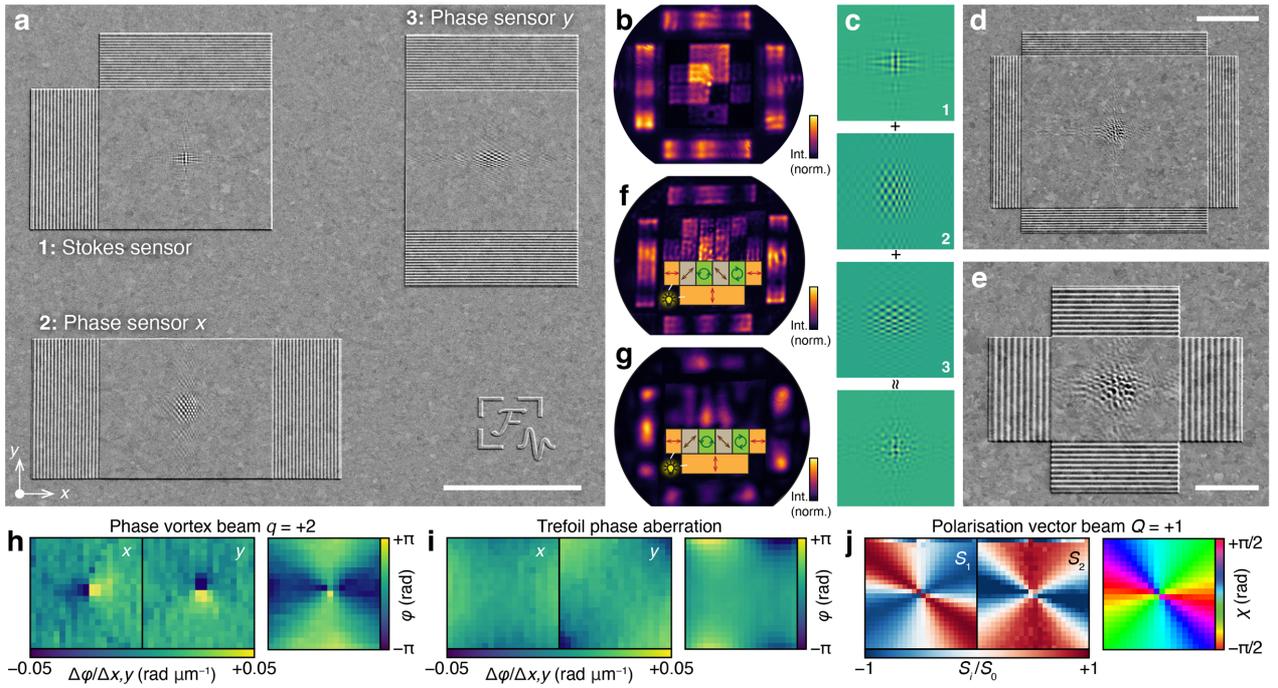

**Figure 4 | Multifunctional Fourier pixels for full sensing of optical fields. a**, SEM (35° tilt) of the first design strategy. Three individual sensors are clustered to detect (1) amplitude and polarisation, (2) phase along $x$, and (3) phase along $y$. They project to different regions in the output (Fourier) plane. **b**, The output from the Fourier pixel in **a**, which is constructed from three different measurements (one for each sensor to correct for microscope misalignments). The central region is from the polarisation sensor (as in Fig. 3g,h). The phase in $x$ and $y$ is seen as interference bands (as in Fig. 3b), placed at the edges of Fourier space. The two redundant bands are at top and bottom for the phase difference in $x$ and left and right for $y$. **c**, Second design strategy for simultaneous sensing of amplitude, phase, and polarisation. Three Fourier elements are superposed to create a single integrated Fourier pixel. **d,e**, SEMs of the experimental implementation of **c** (35° tilt). The sizes of the Fourier elements are 30×30 and 10×10 µm$^2$, respectively. **f,g**, The output Fourier plane from the pixels in **d** and **e**, respectively, which are constructed from three different measurements (one for each sensor functionality). In Extended Data Fig. 8, we show single-shot measurements for the same sensors. The polarisation and phase sensors project in the centre and edges, as in Fig. 4b. However, the polarisation regions are positioned differently, as shown. **h,i**, Experimental demonstration of phase imaging with the Fourier pixel in **a** to map the phase or polarisation of larger complex field profiles, including an optical vortex beam ($q = +2$; $\varphi = e^{2i\theta}$) and trefoil aberrations ($\varphi = x^3 - 3xy^2$). The phase maps are accurately reconstructed from the phase gradients along $x$ and $y$ (see Methods for phase-extraction procedure). **j**, Measurement of the in-plane polarisation angle $\chi$ of a first-order vector beam through the Stokes parameters $S_1$ and $S_2$. The scale bars in **a,d,e** are 20, 10, and 5 µm, respectively.

## Methods

**Fourier-pixel design**

Each Fourier pixel involves SPP generation, SPP propagation, and SPP diffraction. For simplicity, we use light to generate SPPs with sinusoidal gratings. The SPPs are launched when the SPP wavevector, $k_{\text{SPP}}$, satisfies

$$k_{\text{SPP}} = k_{\parallel} + n g_{\text{m}}, \qquad (1)$$

where $k_{\parallel}$ is the in-plane wavevector of photons incident on the grating, $g_{\text{m}} = 2\pi/\Lambda$ is the grating momentum, $\Lambda$ is the grating period, and $n$ is the diffraction order. The photons have wavevector $\mathbf{k}$ with $|\mathbf{k}| = k = \frac{2\pi}{\lambda}$ with wavelength $\lambda$. If the grating contains multiple spatial frequencies, it can couple photons of different wavelengths simultaneously at the same incident angle, launching SPPs with different $k_{\text{SPP}}$.

In a Fourier pixel the generated SPPs propagate in $x$ across the $x, y$ interfacial plane with transverse-magnetic polarisation. We treat the SPPs as scalar reference waves of the form

$$r(x, y) = e^{i k_{\text{SPP}} x}. \qquad (2)$$

The SPPs then encounter the Fourier element that creates a desired complex-valued optical wavefront $g(x, y)$ at a specific output plane via diffraction. The inverse-design process for our Fourier pixels must predict the height profile $h_{\text{p}}(x, y)$ of the Fourier element that generates the desired $g(x, y)$.

In general, diffraction of light by metallic surfaces can be treated by considering the local optical path differences introduced by the structured interface. We apply a similar diffraction model to describe the interaction of SPPs with our Fourier elements. For amplitude and phase, we apply a scalar diffraction model. After the SPP reference wave $r(x, y)$ interacts with the Fourier element, the optical wavefront at the diffractive surface, $f(x, y)$, is described by the relation

$$f(x,y) = r(x,y)\,t(x,y)\,, \tag{3}$$

where we have introduced a complex-valued transparency function $t(x,y)$. It describes how the Fourier element converts the SPPs into the desired wavefront at the diffractive surface. This wavefront then propagates to generate $g(x,y)$. Equation (3) can be rearranged to give

$$t(x,y) = \frac{f(x,y)}{r(x,y)} = f(x,y)\mathrm{e}^{-\mathrm{i}k_{\mathrm{SPP}}x}\,. \tag{4}$$

Our Fourier elements predominantly affect the local phase, $\phi(x,y)$, which is a real quantity. Thus, we must consider experimentally realisable transparency functions of the form $\tilde{t}(x,y) = \mathrm{e}^{\mathrm{i}\phi(x,y)}$. Further, for shallow profiles (and correspondingly small phase shifts) we can linearise the optical response as

$$\tilde{t}(x,y) = \mathrm{e}^{\mathrm{i}\phi(x,y)} \approx 1 + \mathrm{i}\phi(x,y)\,. \tag{5}$$

In this case, $\mathrm{Im}\{\tilde{t}(x,y)\} = \phi(x,y)$, which is a tuneable experimental parameter directly proportional to the local height profile, $h_{\mathrm{p}}(x,y)$. Specifically, $\phi(x,y) = kh_{\mathrm{p}}(x,y)$ in the limit of shallow profiles. Consequently, by adjusting $\phi(x,y)$ we can set the imaginary part of the realisable transparency function $\tilde{t}(x,y)$ equal to that of the desired transparency function $t(x,y)$ (that is, set $\mathrm{Im}\{\tilde{t}(x,y)\} = \mathrm{Im}\{t(x,y)\}$). From equation (4), we then obtain

$$\mathrm{Im}\{\tilde{t}(x,y)\} = \frac{1}{2\mathrm{i}}\left[f(x,y)\mathrm{e}^{-\mathrm{i}k_{\mathrm{SPP}}x} - f^{*}(x,y)\mathrm{e}^{\mathrm{i}k_{\mathrm{SPP}}x}\right]. \tag{6}$$

Plugging our realisable transparency function $\tilde{t}(x,y)$ into equation (3), we then obtain the realisable output $\tilde{f}(x,y)$ from equations (5) and (6)

$$\tilde{f}(x,y) = r(x,y)\,\tilde{t}(x,y) = \mathrm{e}^{\mathrm{i}k_{\mathrm{SPP}}x}\left\{1 + \frac{1}{2}\left[f(x,y)\mathrm{e}^{-\mathrm{i}k_{\mathrm{SPP}}x} - f^{*}(x,y)\mathrm{e}^{\mathrm{i}k_{\mathrm{SPP}}x}\right]\right\} \tag{7}$$

$$\tilde{f}(x,y) = \mathrm{e}^{\mathrm{i}k_{\mathrm{SPP}}x} + \frac{1}{2}\left[f(x,y) - f^{*}(x,y)\mathrm{e}^{2\mathrm{i}k_{\mathrm{SPP}}x}\right]. \tag{8}$$

The realisable output in equation (8) contains three terms: (i) the reference SPP, (ii) the desired output $f(x,y)$, and (iii) an SPP with doubled wavevector. Only the desired output $f(x,y)$ generates an optical wavefront. The other two are dark plasmonic modes. Thus, the realisable Fourier element can create any desired $f(x,y)$. If the required $f(x,y)$ is known, the necessary height profile can be determined from equation (6) in the limit of shallow profiles

$$h_\mathrm{p}(x,y) = \frac{1}{2ik}\left[f(x,y)e^{-ik_\mathrm{SPP}x} - f^*(x,y)e^{ik_\mathrm{SPP}x}\right].\qquad(9)$$

The inverse-design process now involves predicting the complex-valued optical wavefront $f(x,y)$ at the sample plane (at $z=0$) that leads to the desired output $g(x,y)$. For outputs at arbitrary planes ($z=d$), we use the Fresnel approximation. For outputs in the far field (or in our case at $z=2f_\ell$, the back focal plane of a lens with focal length $f_\ell$), we employ the Fraunhofer approximation. The desired output $g(x,y)$ is backpropagated via the angular spectrum method or the inverse Fourier transform, respectively, to obtain the corresponding complex-valued wavefront $f(x,y)$ at the sample plane. In the Fresnel regime, this gives a convolution integral over positional coordinates $(\xi,\eta)$

$$f(x,y) = \iint_{-\infty}^{\infty} g(\xi,\eta)h(x-\xi,y-\eta,-d)d\xi d\eta\,,\qquad(10)$$

where the convolution kernel $h(x,y,z)$ is the Fresnel impulse response

$$h(x,y,d) = \frac{e^{ikd}}{i\lambda d}\exp\left[\frac{i\pi}{\lambda d}(x^2+y^2)\right].\qquad(11)$$

In the Fraunhofer regime, $g(x,y)$ can be represented as function of spatial frequencies $k_x$ and $k_y$

$$\tilde{g}(k_x,k_y) := g\left(x=\frac{k_x f_\ell}{k}, y=\frac{k_y f_\ell}{k}\right),\qquad(12)$$

leading to

$$f(x,y) = i\lambda f_\ell \exp(-i2kf_\ell)\,\mathcal{F}^{-1}\{\tilde{g}(k_x,k_y)\}\,,\text{ or}\qquad(13)$$

$$f(x,y) = \frac{i\lambda f_\ell}{4\pi^2}\exp(-i2kf_\ell)\iint_{-\infty}^{\infty}\tilde{g}(k_x,k_y)\exp[i(k_x x + k_y y)]\,dk_x dk_y\,.\qquad(14)$$

With the formulas for $f(x,y)$ [equations (10) or (14)], we can then determine the required height profile for the Fourier element using equation (9).

The above scalar diffraction model can be generalised to include polarisation and treat vectorial Fourier pixels. SPPs are coherently launched along two perpendicular directions $x$ and $y$. We define two-dimensional vectors for the desired optical output $\mathbf{g}(x,y)$, the backpropagated counterpart $\mathbf{f}(x,y)$, and the reference waves $\mathbf{r}(x,y)$

$$\mathbf{g}(x,y) = \begin{bmatrix} g_x(x,y) \\ g_y(x,y) \end{bmatrix}, \qquad \mathbf{f}(x,y) = \begin{bmatrix} f_x(x,y) \\ f_y(x,y) \end{bmatrix}, \qquad \mathbf{r}(x,y) = \begin{bmatrix} e^{ik_{\text{SPP}}x} \\ e^{ik_{\text{SPP}}y} \end{bmatrix}. \qquad (15)$$

The realisable system response is described by the transparency matrix

$$\widetilde{\mathbf{T}}(x,y) = \begin{bmatrix} \tilde{t}(x,y) & 0 \\ 0 & \tilde{t}(x,y) \end{bmatrix}. \qquad (16)$$

with

$$\tilde{t}(x,y) = 1 + \tfrac{1}{2}\left[f_x(x,y)e^{-ik_{\text{SPP}}x} - f_x^*(x,y)e^{ik_{\text{SPP}}x} + f_y(x,y)e^{-ik_{\text{SPP}}y} - f_y^*(x,y)e^{+ik_{\text{SPP}}y}\right] \qquad (17)$$

and

$$h_{\text{p}}(x,y) = \tfrac{1}{2ik}\left[f_x(x,y)e^{-ik_{\text{SPP}}x} - f_x^*(x,y)e^{ik_{\text{SPP}}x} + f_y(x,y)e^{-ik_{\text{SPP}}x} - f_y^*(x,y)e^{ik_{\text{SPP}}x}\right], \qquad (18)$$

in the limit of shallow profiles. The realisable response at the sample plane is given by $\tilde{\mathbf{f}}(x,y) = \widetilde{\mathbf{T}}(x,y)\mathbf{r}(x,y)$. Each polarisation component contributes five distinct terms. As in equation (8), only the term containing the desired wavefront couples to free space.

**Efficiency of Fourier pixels**

The efficiency of the Fourier pixels was estimated using focusing elements (similar to that in Fig. 1i). The optical power that was focused to the desired spot $I_{\text{focus}}$ was compared to the total power $I_{\text{tot}}$ incident on the input grating. An excitation mask selectively illuminated the grating region under oblique incidence, launching SPPs selectively in the direction of the Fourier element. This element was designed to act as a sinusoidal Fresnel lens. The height profile followed the function

$$h_{\text{p}}(x,y) = A\left\{1 - \exp\left[-\alpha\left(x + \tfrac{L}{2}\right)\right]\right\} \cos\left[\tfrac{\pi(x^2+y^2)}{f_l \lambda} - k_{\text{SPP}}\, x\right], \qquad (19)$$

where $A$ is the height amplitude and $L$ is the side length of the Fourier element in $(x,y)$. The centre of the profile was at $(0,0)$. To supress back reflection and outscattering of the SPPs at the boundary between the grating and the Fourier element, equation (19) includes apodization with $\alpha = k_{\text{SPP}}/5$. The total incident intensity $I_{\text{tot}}$ was obtained from a reflection measurement from flat silver, while the reflected intensity from the grating $I_{\text{grating}}$ gave the incoupling efficiency $\eta_{\text{in}} = 1 - I_{\text{grating}}/I_{\text{tot}}$. The overall efficiency was determined from the integrated focused intensity at 25 µm above the surface as $\eta_{\text{tot}} = I_{\text{focus}}/I_{\text{tot}}$. For blue ($\lambda = 450$ nm, grating amplitude $A = 20$ nm), green (520 nm, $A = 25$ nm), and red (630 nm, $A = 30$ nm) light, the efficiencies $\{\eta_{\text{in}}, \eta_{\text{tot}}\}$ were $\{76.7\%, 23.8\%\}$, $\{72.6\%, 42.2\%\}$, and $\{70.2\%, 41.4\%\}$, respectively. The efficiency for 450 nm light is significantly lower due to increased plasmonic losses in silver at shorter wavelengths.

**Stokes polarimetry**

Arbitrary incoming light fields can be separated into their polarisation components

$$A_x = a_x e^{i\varphi_x}, A_y = a_y e^{i\varphi_y}, \tag{20}$$

with $a$ and $\varphi$ the amplitude and phase, respectively. The Stokes parameters describe the polarisation state of light, defined by

$$S_0 = |A_x|^2 + |A_y|^2 = a_x^2 + a_y^2, \tag{21}$$

$$S_1 = |A_x|^2 - |A_y|^2 = a_x^2 - a_y^2, \tag{22}$$

$$S_2 = 2\,\text{Re}\{A_x^* A_y\} = 2 a_x a_y \cos(\varphi_y - \varphi_x), \tag{23}$$

$$S_3 = 2\,\text{Im}\{A_x^* A_y\} = 2 a_x a_y \sin(\varphi_y - \varphi_x). \tag{24}$$

We assume a uniform polarisation state over the Fourier-pixel area. For detection, we use a linear polariser along the diagonal, 45° from both $x$ and $y$ axes. We measure the intensities in Fourier space corresponding to the isolated $x$ and $y$ components of the incident field, that is, $I_{x,y} = a_{x,y}^2/2$. From these quantities we retrieve the first two Stokes parameters

$$S_0 = 2(I_x + I_y), \tag{25}$$

$$S_1 = 2(I_x - I_y). \tag{26}$$

For the other two Stokes parameters, we recombine the incoupled light fields with tailored phase shifts $\varphi_\text{out}$ between the $x$ and $y$ direction. In general, the total intensity becomes

$$I(\varphi_\text{out}) = \frac{1}{2}\left|A_x + A_y e^{i\varphi_\text{out}}\right|^2 . \tag{27}$$

By choosing the phases $\varphi_\text{out} = \{0, \pi/2, \pi, 3\pi/2\}$, we isolate the remaining Stokes parameters

$$S_2 = I(0) - I(\pi) , \tag{28}$$

$$S_3 = I(3\pi/2) - I(\pi/2) . \tag{29}$$

**Extracting large phase profiles from phase-gradient maps**

The phase profile $\varphi$ was reconstructed from discrete phase-step maps $m_x$ and $m_y$, which represent the local-phase differences between adjacent sampling points. To recover a globally consistent phase, a self-consistency equation between the phase profile and its gradients was solved

$$\nabla^2 \varphi = \frac{\partial m_x}{\partial x} + \frac{\partial m_y}{\partial y} := M , \tag{30}$$

where the middle of equation (30) is obtained from finite differences of the measured phase-step maps. This self-consistency procedure avoids the accumulation of noise that occurs when the phase is retrieved by direct integration. The equation was converted to the spatial-frequency domain using a two-dimensional fast Fourier transform (FFT). In the discrete Fourier domain (denoted by the ^ symbol), the solution is

$$\hat{\varphi} = \frac{\hat{M}}{2\cos(k_x) + 2\cos(k_y) - 4} , \tag{31}$$

with the zero-frequency component set to zero to fix the arbitrary phase offset. An inverse FFT then yielded the phase profile $\varphi$.

**Fourier-pixel fabrication**

The Fourier pixels were fabricated using thermal scanning-probe lithography (TSPL)[16]. PPA [poly(phthalaldehyde)] was used as the thermally sensitive resist. Si (100) wafers (2-inch diameter, 1-mm thickness, Silicon Materials) were first cleaned with oxygen plasma (GIGAbatch, PVA TePla) at 600 W for 2 min. 400 µL of a 12 wt% solution of PPA (Allresist) in anisole (AR 600-02, Allresist) was then spin-coated onto the wafers using a two-step procedure: (i) 5 s at 500 rpm with a ramp of 500 rpm·s$^{-1}$ and (ii) 40 s at 2000 rpm with 2000 rpm·s$^{-1}$. After baking the PPA layer on a hot plate at 110 °C for 2 min, a uniform film thickness of 350–400 nm was obtained. To pattern the PPA, the Fourier-pixel height profiles were loaded into the TSPL tool (NanoFrazor Explore, Heidelberg Instruments Nano). Specific parameters (depth range, pixel size, and feedback gains) were adjusted depending on the design. The writing depth was controlled via electrostatic actuation between the cantilever and substrate. For feedback during writing, the TSPL tool used in-situ topography data.

To replicate the obtained pattern in a plasmonic material, a layer of Ag (>600 nm thick) was deposited onto the patterned PPA via thermal evaporation (Nano 36, Kurt J. Lesker) using silver pellets (99.999%, 1/4-inch-diameter × 1/4-inch-long, Kurt J. Lesker). The deposition rate was maintained at 2.5 nm·s$^{-1}$ under high vacuum (3×10$^{-7}$ mbar)[37]. Subsequently, we attached a 1-mm-thick glass slide (Paul Marienfeld) on top of the Ag surface using ultraviolet-curable epoxy (OG142-95, Epoxy Technology). The glass slide was allowed to rest on the epoxy for 5 min prior to exposure to the ultraviolet light (2 h), which minimised PPA contamination on the Ag. The cured glass/epoxy/Ag stack was stripped from the PPA substrate using a razor blade[13], revealing the surface structure in Ag (inverted from the original PPA pattern). A final cleaning step in anisole (AR 600-02, Allresist) for 2 min removed any residual PPA on the Ag.

**Optical measurements**

The optical setup is depicted in Extended Data Fig. 2. Fourier pixels were measured using a home-built optical setup based on an inverted optical microscope (Eclipse, Ti-U Nikon) equipped with a 50× air objective (TU Plan Fluor, Nikon; numerical aperture of 0.8). A filtered (LLTF Contrast, NKT Photonics; 420–1000 nm accessible wavelengths, linewidth of ~1.5 nm) supercontinuum laser (SuperK FIU-15, NKT Photonics) was used to illuminate the sample at different wavelengths after passing through a short-pass optical filter (FESH0750, Thorlabs). The collimated laser was focused onto the back focal plane of the microscope objective by a 400-mm defocusing lens (L2), resulting in a 100× demagnified Gaussian illumination spot on the sample. Laser-cut-cardboard masks were placed in the excitation path to selectively illuminate the incoupling gratings. A rotatable linear polariser in the excitation path controlled SPP launching from the two orthogonal incoupling gratings (Fig. 2a). More specifically, SPP launching was maximised for one grating by aligning the incoming polarisation with its corrugations or evenly divided over two orthogonal gratings by orienting the polarisation at 45° between them. Diffracted light off the substrate was passed through a circular aperture in the image plane to isolate light emanating from the Fourier element. For all Fourier pixels designed to project the output to the far field (Extended Data Fig. 2a), the back focal plane of the microscope objective was imaged on a digital camera (Zyla PLUS sCMOS, Andor). For Fourier pixels that project to arbitrary planes (Extended Data Fig. 2b), the lens L6 was removed to image the sample plane or at planes displaced along the optical axis. The exact plane could be selected by moving the microscope turret relative to the sample.

For measurements with incoherent illumination, the laser source was replaced by a halogen lamp spectrally filtered by a 10 nm full-width-at-half-maximum (FWHM) bandpass filter. The angular spectrum of the lamp was further restricted by placing a mask in a Fourier plane in the excitation path, ensuring SPP launching into a single direction.

To control the phase of the incoming laser, a liquid-crystal spatial-light modulator (SLM; Pluto NIR011, Holoeye) was used. The SLM display was imaged onto the Fourier pixel using a 750-mm lens (instead of the 400-mm lens L2) and the microscope objective, achieving a demagnification of 187.5×. The SLM display was subdivided into two regions, and the relative phase between them was systematically varied between $0$ and $2\pi$. The two opposing incoupling gratings (Fig. 3a) were illuminated by the two regions of the SLM, thereby tuning the incoming phase difference. This measurement procedure was also used to systematically project parts of large complex phase profiles onto the phase sensors of the Fourier pixel (Fig. 4h,i).

To systematically control the in-plane polarisation of the incoming laser (Fig. 4j), a rotatable mechanical mount (PRM1Z8, Thorlabs) was used in combination with a linear polariser and custom-built control software. This configuration provided adjustable linear polarisation, as the laser source is unpolarised in-plane. A quarter-wave plate was used together with the linear polariser, oriented at −45° or +45°, to generate left- or right-circularly polarised light (Fig. 3i,j), respectively. In the detection path, a linear polariser projected the diffracted light onto the axis at 45° between the gratings to extract the Stokes parameters.

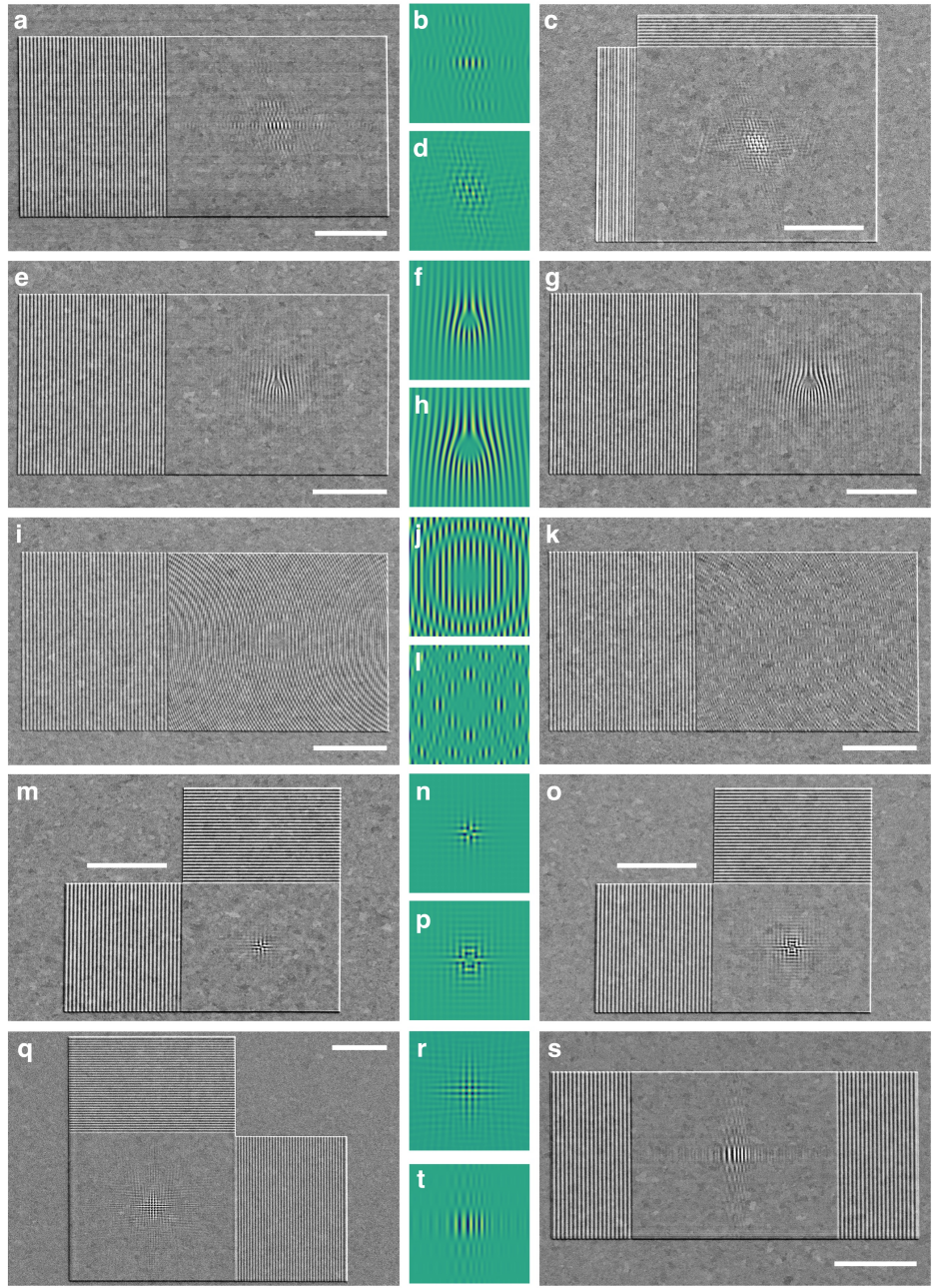

**Extended Data Figure 1 | Electron micrographs and designs of Fourier pixels that shape and sense light.** Scanning electron micrographs (SEMs) of Fourier pixels (35° tilt) for: **a**,**b**, a single-color ETH logo (Fig. 1g); **c**,**d**, a multi-colour ETH logo (Fig. 1h); **e**–**h**, vortex beams with topological charge $q$ of +3 and +5 (Fig. 1e,f); **i**,**j**, a diffraction-limited focal spot (Fig. 1i); **k**,**l**, a grid of foci (Fig. 1j); **m**–**p**, vector beams of orders $Q = +1$ and +2 (Fig. 2c,d); **q**,**r**, multiplexed images of arrows (Fig. 2e,f); **s**,**t**, phase sensing that diffracts light to the centre of Fourier space (Fig. 3b). All scale bars are 10 µm. The design (height profile) is shown for the central 10×10 µm² of the Fourier element.

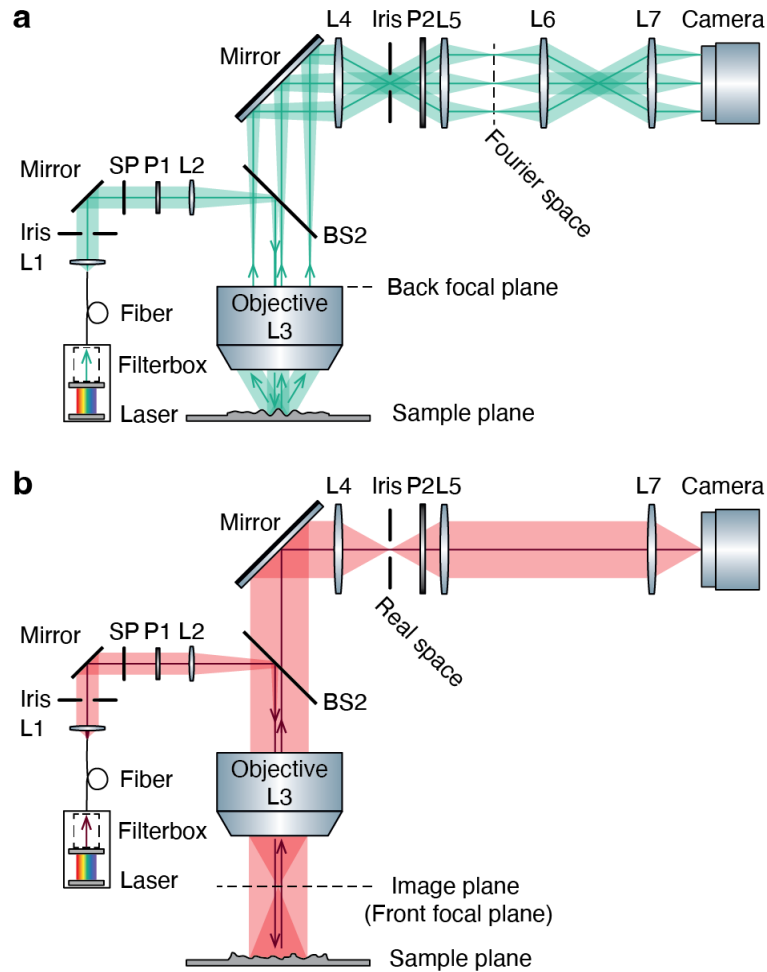

**Extended Data Figure 2 | Optical setup for measuring Fourier pixels. a**, Schematic of the optical setup (Fourier microscope) to measure the output of a Fourier pixel that projects its output to Fourier space (the back focal plane of our microscope objective). **b**, Schematic of the optical setup to measure the output of a Fourier pixel that projects to an arbitrary image plane above the sample. The setup is also used to image the sample plane (for example, Fig. 1i,j). Lens L6 from **a** is removed to project spatial information on the camera instead of angular information. Both setup configurations include lenses (L), polarizers (P), short-pass filters (SP), and beam splitters (BS). See Methods for further details.

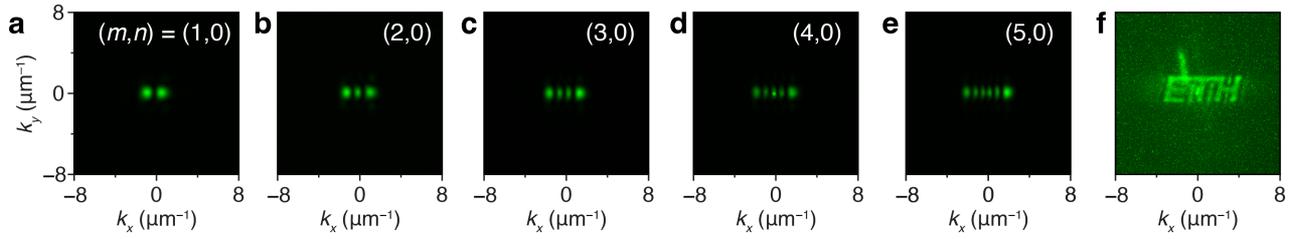

**Extended Data Figure 3 | Hermite–Gaussian beams and generation of images using incoherent light. a–e**, Constructed Hermite–Gaussian beams of orders $(m, n) = (x, 0)$ with $x = 1\text{--}5$, obtained from a Fourier pixel operating at 565 nm. Panels **a–e** correspond to increasing beam order. **f**, Generation of the 'ETH' logo using 546-nm incoherent light from a halogen lamp.

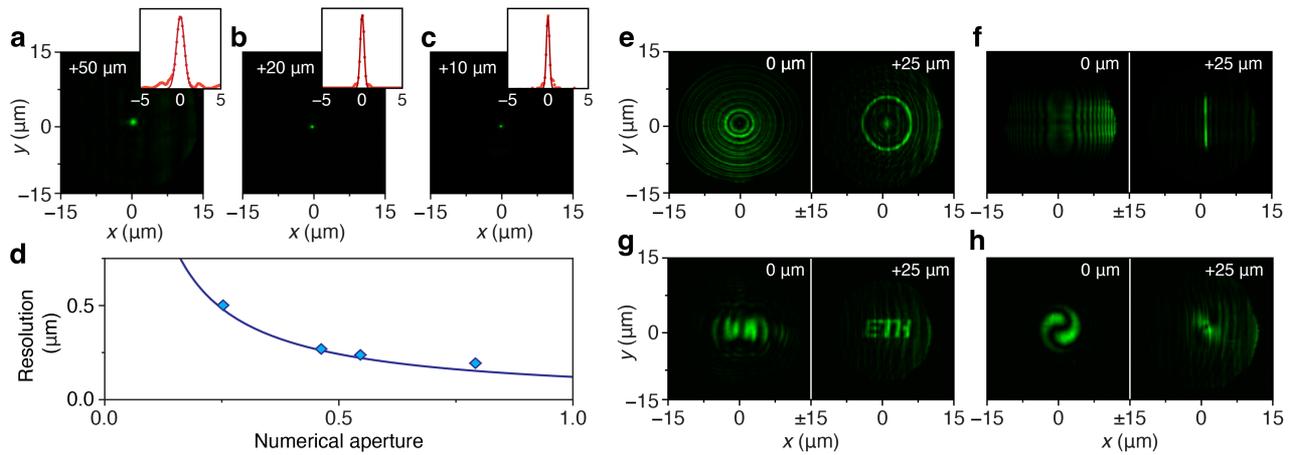

**Extended Data Figure 4 | Fourier pixels that project to desired output planes. a**, Optical output of a Fourier pixel that focuses with a focal length $f_\ell$ of 50 μm. Inset: crosscut through the centre of the focal plane, showing a diffraction-limited spot with a standard deviation $\sigma$ of 0.5 μm. **b,c**, Same as **a**, but for Fourier pixels with focal distances of 20 and 10 μm, respectively. The foci are significantly tighter, as expected. **d**, Resolution of the Fourier-pixel lens (defined as the standard deviation of the focal spot) as a function of its numerical aperture (NA), given by $L/(2f_\ell\sqrt{1 + (L/2 f_\ell)^2})$, where $L$ is the side length of the lens structure. We use effectively only 80% of the actual side length ($L = 24$ μm) of the lens because of apertures in the detection path. The data points overlap well with the expected resolution from diffraction, that is, the Airy formula, $0.218\lambda/\text{NA}$, at the operating wavelength of 550 nm. **e–h**, Fourier pixels with a focal length of 25 μm that project a circle, a line, an 'ETH' logo, and a Gaussian vortex beam ($q = +1$), respectively. For each, the left image shows the sample in focus, while the right shows 25 μm above the sample plane.

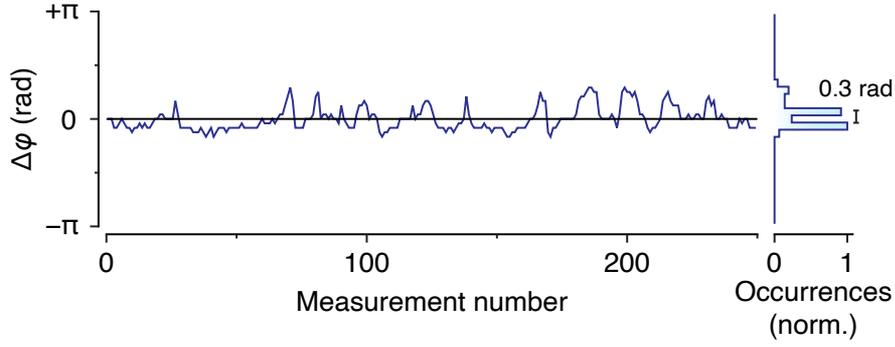

**Extended Data Figure 5 | Instability of our measurement setup visualised by our Fourier phase sensor.** Fluctuations $\Delta\varphi$ in the phase readout at fixed input phase by the Fourier pixel over 250 consecutive measurements reveals a standard deviation of 0.3 rad. These variations translate to incoming-angle fluctuations of 0.07°, considering the pixel size of 25 µm and the wavelength of 550 nm. We attribute these changes to instability in our measurement setup.

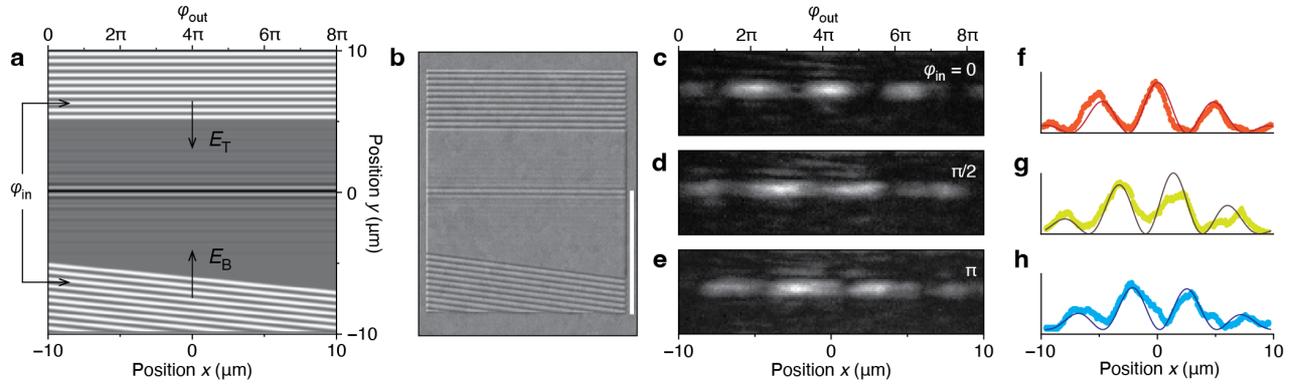

**Extended Data Figure 6 | Fourier pixel for in-plane phase sensing. a**, Height profile of a Fourier-pixel design that measures phase at the metal interface. Two opposing gratings (top and bottom) launch surface plasmon polaritons (SPPs) towards the Fourier element (middle). The bottom grating is tilted to add a position-dependent phase to the SPPs. The electric fields from the top $E_T = e^{ik_{SPP}y}$ and bottom $E_B = e^{-ik_{SPP}y}e^{i\varphi_{in}}e^{i\varphi_{out}(x)}$ then constructively and destructively interfere in the centre of the device, depending on the phase difference of the incoming light $\varphi_{in}$ and the phase difference induced by the phase-shifted incoupling grating $\varphi_{out}$. From the positions of the dips and peaks, the phase difference of the incoming light can be retrieved. **b**, SEM of the in-plane phase sensor in silver (35° tilt). Scale bar is 20 µm. **c**, Outscattered intensity of 532-nm light for $\varphi_{in} = 0$. SPPs constructively interfere when $\varphi_{out} = 2N\pi$, with $N$ an integer. **d,e**, Same as **c**, but for different $\varphi_{in}$. The intensity maxima shift to spatial positions where $\varphi_{out}$ counteracts $\varphi_{in}$. **f–g**, Data points: crosscuts through **c–e**. Solid lines: product of the SPP intensity at the centre of the Fourier element $|E_T(x, y = 0) + E_B(x, y = 0)|^2$ and a Gaussian envelope reproducing the width of the laser beam.

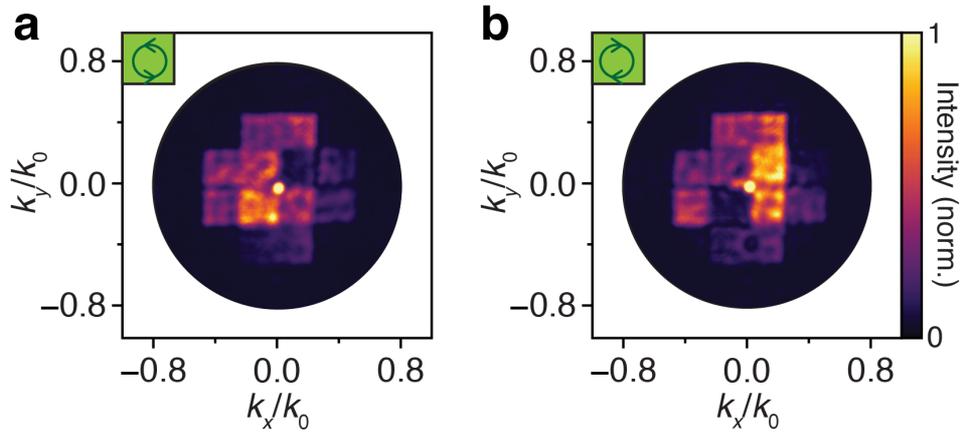

**Extended Data Figure 7 | Polarisation sensing. a,b,** Measured optical output as in Fig. 3g,h but for optical inputs of left-circular and right-circular polarisation.

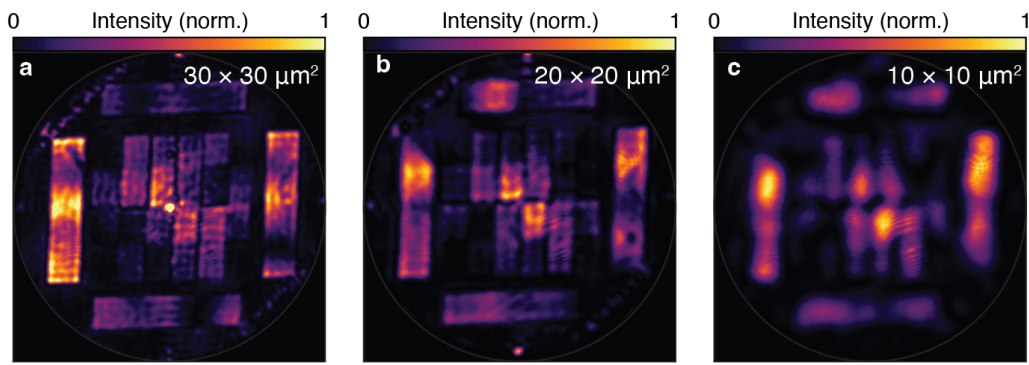

**Extended Data Figure 8 | Single-shot full-field sensing.** Single-shot full-field characterisation using the **a**, 30×30 µm², **b**, 20×20 µm², and **c**, 10×10 µm² multifunctional Fourier elements, introduced in Fig. 4d,e. The optical features are broadened in Fourier space for the smaller structures because of finite-size effects in diffraction. The diagonal input polarisation can be retrieved by all sensing pixels, but the phase readout is slightly distorted. We speculate that the linear polarisation of the diffracted fields is distorted by our optical microscope which hinders the observation of destructive/constructive interference in our phase measurements.